\documentclass[a4paper,fleqn]{cas-dc}

\usepackage{relsize}
\usepackage{titlesec} 
\usepackage{cuted}  

\titleformat*{\section}{\small\bfseries}
\titleformat*{\subsection}{\small\bfseries}
\titleformat*{\subsubsection}{\small\bfseries}

\usepackage[numbers,sort&compress,longnamesfirst]{natbib}
\usepackage{xcolor}  
\usepackage{soul}
\definecolor{authorcolor}{rgb}{0,0,0}  
\definecolor{myblue}{RGB}{12, 157, 201}
\definecolor{myblue1}{RGB}{255, 255, 255}
\usepackage{url}
\usepackage{amsmath,amsfonts}
\usepackage{hyperref}
\hypersetup{
    colorlinks=true, 
    linkcolor=myblue,  
    citecolor=myblue,  
    urlcolor=myblue  
}
\usepackage{newtxtext} 
\usepackage{caption}
\usepackage{float} %
\usepackage{lipsum} %
\usepackage{tcolorbox} %
\newtcolorbox{mybox}{
    colback=white, %
    colframe=black, %
    boxrule=1pt, %
    arc=0pt, %
    left=8pt, %
    right=8pt, 
    top=6pt, 
    bottom=6pt, 
    fontupper=\rmfamily 
}
\newtcolorbox{myboxsingle}[1][]{
    colback=white,
    colframe=black, %
    boxrule=0.5pt, %
    arc=0pt, %
    left=8pt, %
    right=8pt, %
    top=6pt, %
    bottom=6pt, 
    fontupper=\rmfamily, %
    #1
}

\usepackage[ruled]{algorithm2e}

\usepackage{fancyhdr}
\def\tsc#1{\csdef{#1}{\textsc{\lowercase{#1}}\xspace}}
\tsc{WGM}
\tsc{QE}
\tsc{EP}
\tsc{PMS}
\tsc{BEC}
\tsc{DE}

\begin{document}
\let\WriteBookmarks\relax
\def\floatpagepagefraction{1}
\def\textpagefraction{.001}


\title [mode = title]{Distributed Reinforcement Learning Using Local Smart Meter Data for Voltage Regulation in Distribution Networks}

\author[1]{\textcolor{authorcolor}{Dong Liu}}
\credit{Conceptualization, Methodology, Software, Validation, Writing - original draft}

\author[2]{\textcolor{authorcolor}{Juan S. Giraldo}}
\credit{Conceptualization, Writing - review \& editing}

\author[1]{\textcolor{authorcolor}{Peter Palensky}}
\credit{Funding acquisition}

\author[1]{\textcolor{authorcolor}{Pedro P. Vergara}}[orcid=0000-0003-0852-0169]
\cormark[1]
\ead{P.P.VergaraBarrios@tudelft.nl}
\credit{Writing - review \& editing, Supervision, Funding acquisition}

\affiliation[1]{organization={Intelligent Electrical Power Grids Group},
    addressline={Delft University of Technology}, 
    postcode={2628CD}, 
    country={The Netherlands}}
 
\affiliation[2]{organization={Techno-Economic Energy Transition Studies, Netherlands Organisation for Applied Scientific Research},
    postcode={2595 DA}, 
    country={The Netherlands}}


\cortext[cor1]{Corresponding author}

\begin{abstract}
Centralised reinforcement learning (RL) for voltage magnitude regulation in distribution networks typically involves numerous agent-environment interactions and power flow (PF) calculations, inducing computational overhead and privacy concerns over shared data. Thus, we propose a distributed RL algorithm to regulate voltage magnitude. First, a dynamic Thevenin equivalent model is integrated within smart meters (SM), enabling local voltage magnitude estimation using local SM data for RL agent training, and mitigating the dependency of synchronised data collection and centralised PF calculations. To mitigate estimation errors induced by Thevenin model inaccuracies, a voltage magnitude correction strategy that combines piecewise functions and neural networks is introduced. The piecewise function corrects the large errors of estimated voltage magnitude, while a neural network mimics the grid's sensitivity to control actions, improving action adjustment precision. Second, a coordination strategy is proposed to refine local RL agent actions online, preventing voltage magnitude violations induced by excessive actions from multiple independently trained agents. Case studies on energy storage systems validate the feasibility and effectiveness of the proposed approach, demonstrating its potential to improve voltage regulation in distribution networks.

\vspace{-\baselineskip}
\end{abstract}


\maketitle

\section{Introduction}
\subsection{Motivation}
With the popularization of smart electrical appliances and the integration of distributed energy resources (DERs), energy demand in distribution networks (DNs) has become increasingly uncertain, presenting challenges to voltage management. Specifically, the uncertainty and volatility of flexible assets likely lead to more voltage magnitude violations~\textcolor{myblue}{\cite{olivier2015active, wang2021deep}}. Medium-voltage distribution networks (MVDNs) exhibit a greater dispatchable load proportion compared to low-voltage distribution networks (LVDNs), thereby facilitating centralized or distributed voltage regulation based on demand response. However, dispatchable demand is of a small size in LVDNs. Existing centralized and distributed approaches rely on prior topology information or synchronous interaction of distributed agents during training, leading to privacy issues and delayed training~\textcolor{myblue}{\cite{liu2025model, mylonas2025safe}}. Additionally, accurate estimation of voltage magnitudes at the smart meter (SM) level using only local measurements remains a challenging task. Voltage magnitude violations caused by adjusted household appliances may be fined by the distribution system operator (DSO) in the future to guarantee network steady-state security during serious line congestion or high demand periods \textcolor{myblue}{\cite{arghavani2017unbalanced}}.

\subsection{Literature review}
Traditional centralized voltage magnitude regulation approaches rely on collected datasets and coordination among the multiple resources~\textcolor{myblue}{\cite{10242604,10121692}}, e,g.,  on-load-tap changer (OLTC), capacitor banks~(CB), photovoltaic (PV) systems, etc. A coordination strategy was proposed in \textcolor{myblue}{\cite{7062068}} to fully use DERs to provide auxiliary services in a more timely manner than OLTC while reducing the utilization of other devices to efficiently maintain the steady-state security of DNs and the lifespan of the devices. A distributed active management approach was introduced in \textcolor{myblue}{\cite{7063267}} to regulate the voltage magnitude by monitoring the output power of the PV system while reducing the communication frequency among the controllers in DNs. A model predictive control (MPC)-based coordinated method was proposed in \textcolor{myblue}{\cite{10287147}} based on the approximate surrogate model of DNs to dispatch fast and slow response assets, aiming to efficiently avoid voltage magnitude violations. To pro-actively manage voltage magnitude, a mixed integer linear programming model is proposed in \textcolor{myblue}{\cite{10026490}} to estimate the export limit of household net power through a centralized model. Nevertheless, approaches relying on resources in DNs, multiple objectives need to be traded off, e.g., the benefits of~OLTC and the reduction of the transformer lifespan. Additionally, the computational complexity of solving the optimization problem increases significantly with system scale, limiting the applicability of the MPC-based and mixed integer nonlinear programming models to scenarios with smaller time intervals. Furthermore, most centralized coordinated approaches require the network topology or a surrogate model, and synchronised measurement collection, which brings privacy risks for customers.

As an efficient dispatching resource, energy storage systems~(ESSs) and electric vehicles (EVs) are gradually being incorporated into the active management of DNs, especially in peak demand regulation and voltage regulation. These approaches share the characteristic of enabling the flexible dispatch of energy storage equipments while the integration of EVs as auxiliary devices for voltage magnitude regulation introduces greater uncertainty~\textcolor{myblue}{\cite{9732531}}. An MPC-based distributed active management strategy was introduced to enable EVs to consume the extra power of the PV system after voltage magnitude regulation, avoiding unreasonable PV curtailment~\textcolor{myblue}{\cite{9726706}}. In~\textcolor{myblue}{\cite{7752949}}, an adaptive decentralized strategy was constructed to simultaneously mitigate line congestion and voltage magnitude violations. To efficiently exploit the ability to regulate multiple resources in DNs, a distributed control approach was introduced in~\textcolor{myblue}{\cite{kang2024distributed}} for virtual ESSs to mitigate the negative impact on voltage magnitude regulation induced by communication delays. A voltage estimation approach with chance constraints was introduced in~\textcolor{myblue}{\cite{10757433}} to handle the uncertainty induced by PV, ESSs and EVs to ensure the accuracy of voltage magnitude estimation under dynamic environments and changing topologies, enhancing the robustness of voltage magnitude regulation. To enhance the feasibility of voltage magnitude regulation approaches in practical cases, a distributed MPC-based method was introduced in~\textcolor{myblue}{\cite{9526869}} to mitigate voltage violations in unbalanced DNs by adjusting the reactive power of EVs chargers. With the reduction of ESSs costs, their introduction in DNs is expected to increase gradually, deployed via technology based on batteries or hydrogen storage technology for long-term energy shifting~\textcolor{myblue}{\cite{chatzigeorgiou2024review}}. However, battery dispatching technology used in energy communities is not suitable for household ESSs or EVs dispatching, because individual loads are more difficult to predict, and many privacy issues are involved~\textcolor{myblue}{\cite{10836703}}. Moreover, model-based approaches mentioned above rely on the DSO to perform centralized PF calculations at each decision-making step, using the network topology and collected system states. In contrast, a trained reinforcement learning (RL)-based agent makes decisions based on the learned policy and observed system state, without the need to solve the PF equations in real time. This significantly reduces the computational burden during the online application and enhances the scalability and responsiveness of the control strategy.

Unlike traditional data-driven approaches, RL does not rely on labelled or forecasting datasets but instead uses a reward mechanism to guide the agent towards maximizing cumulative rewards over time. This approach has proven highly effective in solving complex, long-term decision-making problems across various domains. A two-layer RL-based approach is introduced in \textcolor{myblue}{\cite{9760006}} to coordinate the control of slow-response discrete resources and fast-response continuous resources, effectively preventing voltage violations in DNs. In \textcolor{myblue}{\cite{zhang2025explainable}}, an explainable RL-based approach aims to enhance the practical application of RL in voltage regulation. The trained multi-agent system is capable of coordinating the control of multiple DERs for voltage magnitude regulation, including ESSs and hydrogen storage systems. However, the reliability of the actions of neural network-based RL agents is not yet fully verified. To address this, human expertise is incorporated into the actor-critic algorithm~\textcolor{myblue}{\cite{10335935}}. The resulting agents optimise suspicious actions, thereby ensuring that the PV power does not cause extra voltage violations to DNs. A similar study is presented in~\textcolor{myblue}{\cite{10879343}} where a physics-shielded RL algorithm is deployed to enforce DERs coordination, while ensuring the safety of the equipment. 

To achieve efficient coordinated control of heterogeneous DERs, a voltage-reactive power sensitivity is introduced to guide the convergence direction of the multi-agent training process, thereby reducing the frequency of voltage magnitude violations~\textcolor{myblue}{\cite{9983850}}. Considering the similarity between the appliances in different houses, transfer learning was incorporated in RL-based approaches to enable a basic agent transfer from one home to another without leading to re-training of models and avoiding privacy leakage~\textcolor{myblue}{\cite{9941190}}. To train RL-based approaches under distributed datasets, federated learning was employed to enable agents training without compromising the household's privacy~\textcolor{myblue}{\cite{9247266}}. To enhance the performance of the trained agents under various scenarios, clustering algorithms were used to group the training datasets, which were then assigned to train multiple agents. During online application, scenarios (i.e., state sets) are sent to the agent trained on the same or a similar category to manage the household appliances \textcolor{myblue}{\cite{9716893}}. To address the uncertainty in load profiles resulting from human activities, a recognition model was integrated into the RL-based household energy management systems in \textcolor{myblue}{\cite{wu2024smart}}, facilitating the efficient management of household appliance operations.

Besides their potential, most RL-based voltage magnitude regulation approaches rely on centralized training frameworks and PF models, which introduce several limitations. First, they often overlook practical issues (e.g., communication delays) and raise privacy concerns due to the need for extensive data aggregation~\textcolor{myblue}{\cite{10171153}}. In distributed scenarios employing centralized environments, all SM data must be collected before decision-making, further exacerbating communication overhead and exposing sensitive user information. Second, when agents are trained solely on local SM data without awareness of the operational states of neighbouring nodes, significant uncertainty is introduced into local voltage magnitude regulation. This can result in voltage magnitude violations triggered by synchronized behaviours, e.g., excessive simultaneous charging or discharging across multiple nodes. Third, agents trained for voltage magnitude regulation at a single node exhibit limited effectiveness in practice. Their performance is constrained by technical restrictions, such as the capacity of ESS, making them insufficient for ensuring acceptable voltage profiles across the DNs. Moreover, privacy-preserving coordination mechanisms for RL agents trained in a distributed manner remain an open question. 

\vspace{-0.2cm}
\subsection{Contributions}
To address these research gaps, we developed a distributed RL algorithm for voltage magnitude regulation in LVDNs. The three key contributions of this work are summarized as follows:
\begin{itemize}
    \item A dynamic Thevenin equivalent model is used to construct a local environment, enabling distributed RL agents to asynchronously learn policies based on individual SM data. Compared to centralized approaches, the proposed algorithm eliminates the need for synchronized data collection in central PF calculation, thereby reducing privacy risks and the interaction frequency between agents and the DSO. 

    \item To mitigate errors brought by the inaccuracies of the Thevenin equivalent model, a voltage magnitude correction strategy is developed based on piecewise functions and neural networks. The piecewise function is employed to correct large assignment errors, while the neural network models the DN's sensitivity to the agent's actions, ensuring precise voltage adjustments.

    \item To prevent voltage magnitude violations caused by the excessive simultaneous actions of locally trained RL agents, we propose a coordination strategy that integrates an optimization model with a coordination scaler. This strategy locally refines agent actions, ensuring each ESS operates within safe limits and preventing excessive charging or discharging that could worsen voltage magnitude violations.
\end{itemize}
\vspace{-0.4cm}

\section{Preliminaries}
\subsection{ESSs-based Voltage Magnitude Regulation}
Consider a DN with a set of nodes $\mathcal{N}$, where a subset of nodes $\mathcal{B} \subseteq \mathcal{N}$ are equipped with ESSs. Voltage magnitude regulation approach via ESSs can be performed by dynamically dispatching their power outputs, which can be formulated as a non-linear optimisation problem. The typical objective function of this problem is to use the minimal active power to avoid voltage magnitude violation during the period $\mathcal{T}$ \textcolor{myblue}{\cite{9830781,10648969}}, which is expressed as \textcolor{myblue}{(\ref{eq1})}. The constraints are given by \textcolor{myblue}{(\ref{eq3})}–\textcolor{myblue}{(\ref{eq10})}.
\vspace{-0.1cm}
\begin{equation}
    \label{eq1}
    \min_{P^b_{m,t}} \quad \sum_{t \in \mathcal{T}} \sum_{m \in \mathcal{B}} C_t^b|P^b_{m,t}|
\end{equation}
\textit{\textbf{s.t.}}
\vspace{-0.2cm}
\begin{align}
    \label{eq3}
    &\hspace{-0.6cm} \sum_{km\in {\cal L}} P_{km,t} + P_{m,t}^b - \sum_{mn\in \cal{L}}P_{mn,t}- \sum_{mn\in \cal{L}} \frac{P_{mn,t}^2 + Q_{mn,t}^2}{V_{m,t}^{2}}  \mathrm{R_{mn}} = \mathrm{P_{m,t}^{D}} \quad \nonumber\\
    & \hspace{4.6cm}\forall m\in {\cal N},\forall t\in {\cal T} 
\end{align}
\vspace{-0.5cm}
\begin{align}
    \label{eq4}
    &\hspace{-0.6cm} \sum_{km\in \cal{L}} Q_{km,t} + Q_{m,t}^b - \sum_{mn\in  \cal{L}} Q_{mn,t} - \sum_{mn\in  \cal{L}} \frac{P_{mn,t}^2 + Q_{mn,t}^2}{V_{m,t}^{2}}  \mathrm{X_{mn}}   = \mathrm{Q_{m,t}^{D}} \nonumber\\
    & \hspace{4.6cm} \forall m\in {\cal N},\forall t\in {\cal T} \ \\
     \label{eq5}
    &\hspace{-0.6cm} {V}_{m,t}^{2} - {V}_{n,t}^{2} = 2 (\mathrm{R_{mn}} {{P}_{mn,t}}+ \mathrm{X_{mn}} {{Q}_{mn,t}}) \  \nonumber \\ 
    &\hspace{3.0cm}- \frac{P_{mn,t}^2 + Q_{mn,t}^2}{V_{m,t}^{2}}(\mathrm{R^2_{mn}} + \mathrm{X^2_{mn})} \nonumber \\
    & \hspace{3.1cm} \forall m,n\in {\cal N},\forall mn\in {\cal L},\forall t\in {\cal T} \ \\
    \label{eq6}
    & \hspace{-0.6cm} \underline{\text{V}} \leq V_{m,t} \leq \overline{\text{V}} \hspace{3.6cm} \forall m\in {\cal N},\forall t\in {\cal T} \\
    \label{eq7}
    &\hspace{-0.6cm} \underline{\text{P}}^b \leq P^b_{m,t} \leq \overline{\text{P}}^b, \hspace{3.5cm} \forall m \in \mathcal{B},\forall t\in {\cal T}\\
    \label{eq9}
    &\hspace{-0.6cm} \underline{\text{SOC}} \leq SOC_{m,t} \leq \overline{\text{SOC}}, \hspace{2.358cm} \forall m \in \mathcal{B}, \forall t\in {\cal T}\\
    \label{eq10}
    &\hspace{-0.6cm} SOC_{m,t+1} = SOC_{m,t} + \frac{P^b_m \Delta t}{\text{E}^{rated}_m}, \hspace{1.66cm} \forall m \in \mathcal{B}, \forall t\in {\cal T}
\end{align}
where $C_t^{b}$ denotes the compensation price paid to the ESS owners for participating in voltage magnitude regulation. Indices $m$ and $n$ represent nodes in $\mathcal{N}$, while $mn$ and $km$ denote lines in $\mathcal{L}$.

$V_{m,t}$ is the voltage magnitude at node $m$ at time $t$, which is bounded by the lower limit $\underline{\text{V}}$ and upper limit $\overline{\text{V}}$. The power balance is modelled by constraints \textcolor{myblue}{(\ref{eq3})} and \textcolor{myblue}{(\ref{eq4})}. The fourth term in constraints \textcolor{myblue}{(\ref{eq3})} and \textcolor{myblue}{(\ref{eq4})} represents the active power loss and reactive power loss in the line connecting nodes $n$ and $m$. Constraint \textcolor{myblue}{(\ref{eq5})} formulates the voltage magnitude drop across the lines. Parameters $\mathrm{R_{mn}}$ and $\mathrm{X_{mn}}$ are the resistance and reactance of the line $mn$. Variables $P_{km,t}$ and $P_{mn,t}$ represent the active power flowing into and out of node $m$ at time $t$, respectively. $Q_{km,t}$ and $Q_{mn,t}$ represent the reactive power flowing into and out of node $m$ at time $t$, respectively. 

$P^b_{m,t}$ and $Q^b_{m,t}$ denote the active and reactive power outputs of the ESS at node $m$ and time $t$, respectively. The two variables are assumed to be interrelated via the power factor $\cos\theta$. A positive value indicates charging behavior, whereas a negative value represents discharging behaviour. When a ESS is not installed at $m$, the value is set to 0. The active $P^b_{m,t}$ of the batteries are bounded by their power limits $\overline{\text{P}}^b,\underline{\text{P}}^b$, reads as \textcolor{myblue}{(\ref{eq7})}. $\mathrm{P_{m,t}^{D}}$ and $\mathrm{Q_{m,t}^{D}}$ represent the net demand at node $m$ at time $t$. The State of Charge (SOC) of the ESS must remain within its allowable range and is updated based on the ESS power output, which is expressed as expression \textcolor{myblue}{(\ref{eq9})} and \textcolor{myblue}{(\ref{eq10})}. $E^{rated}_m$ is the capacity of the ESS at node $m$. The ESS charging/discharge efficiency is assumed as 100\%.

\subsection{Centralized RL Algorithm}
RL algorithms enable agents to derive optimal decision-making strategies by maximizing cumulative rewards through interaction with predefined environments. The RL algorithm consists of several key components: agent, environment, state, action, and reward. The agent interacts with the environment by taking actions based on its current state, and the environment provides feedback in the form of rewards. The ultimate goal is to develop an optimal policy that maps states to actions, maximizing long-term rewards.

The dispatching of ESSs can be formulated as a Markov decision process, where the $SOC_{t+1}$ at a given time depends on the previous $SOC_{t}$ and the ESS's charging/discharging power (i.e., actions $a_t$), as represented in \textcolor{myblue}{\eqref{eq7}}. The general deployment process of a single-agent RL for centralized scheduling of multiple ESSs is summarized below~\textcolor{myblue}{\cite{vergara2022optimal, 10214347}}:

\textit{\textbf{Initialization}}
\begin{itemize}
    \item Environment construction: it represents the LVDN, incorporating its components, constraints, and dynamic behavior. Normally, it consists of PF models and related functions.
    \item State definition: the voltage magnitudes of $N$ nodes and $SOC_t$ of the ESS are defined as state variables $S_t=[SOC_t, V_{1,t}, V_{2,t}, ..., V_{N,t}]$.
    \item Reward function. DSOs aim to minimise the active power used for voltage magnitude violation. The reward is formulated as:
    \begin{align}
     &\hspace{-0.2cm}  R_t = -C_t^b\sum_{m\in \mathcal{B}} |P^b_{m,t}| \nonumber \\
     & \hspace{0.5cm}+ \sum_{m\in \mathcal{N}} min\{0, \frac{V_{max} - V_{min}}{2} - |V_{m,t} - V_{max}|\}
    \end{align} 
    The reward $R_t$ increases as the total active power used in voltage magnitude regulation decreases.
    \item The surrogate model and policy $\pi$ in the RL agent are initialized.
\end{itemize}

\vspace{0.15cm}

\textit{\textbf{Agent Observation}}

The agent observes the current state $S_t$, and then selects an action $a_{t+1}$ (i.e., charging / discharging power $P^b_{t}$) based on the policy $\pi$:
\begin{align}
    \hspace{0.99cm} a_{t+1} \sim \pi(a_t | S_t)\\
    \textit{\textbf{s.t.}} \hspace{0.8cm} Constraints\quad \textcolor{myblue}{\eqref{eq7}-\eqref{eq10}} \nonumber
\end{align}
The ESS charges/discharges based on the obtained action $a_{t}$.

\vspace{0.2cm}

\textit{\textbf{Interaction}}

The environment updates the system states by solving the PF model with constraints \textcolor{myblue}{\eqref{eq3}}-\textcolor{myblue}{\eqref{eq6}}. The reward $R_{t}$ is obtained for assessing the action $a_t$. The new state $S_{t+1}$ is obtained in the environment and sent back to the agent.
\begin{align}
    S_{t+1} &= f_{PF}(a_t, S_{t}, P_t^{D}, Z_R, Z_X, SOC_t)
\end{align}
where $f_{PF} $ represents the transition function based on PF calculation. $Z_R$ and $Z_X$ represents the line resistance and reactance parameters, respectively.

\vspace{0.15cm}

\textit{\textbf{Policy $\pi$ updating}}

The proximal policy optimization (PPO) algorithm is employed to update the policy $\pi$. To ensure stable learning and prevent excessively large policy updates, a clipped surrogate objective function is utilized \textcolor{myblue}{\cite{10402062}}.

$\mathcal{Q}(S_t, a_t)$ is the action-value function, representing the expected return when taking action $a_t$ at state $S_t$ and following policy $\pi$, formulated as \textcolor{myblue}{\eqref{eq17}}. $\mathcal{V}(S_t)$ is the state-value function, representing the expected return from state $S_t$ following policy $\pi$, formulated as~\textcolor{myblue}{\eqref{eq18}}.
\begin{align}
    \label{eq17}
    &\mathcal{Q}(S, a) = \mathbb{E}_{\pi} \left[ \sum_{t=0}^{\infty} \gamma^t R_t \mid S_0 = S, a_0 = a \right]\\
    \label{eq18}
    &\mathcal{V}(S_t) = \mathbb{E}_{\pi} \left[ \sum_{t=0}^{\infty} \gamma^t R_t \mid S_0 = S \right]
\end{align}
where $\gamma$ is the discount factor ($ 0 \leq \gamma \leq 1 $), balancing immediate and future rewards.  

The parameters $\theta_{k}$ of policy $\pi_{\theta_{k}}$ are updated using the gradient descent algorithm, and the value function is updated using the mean-squared error loss.
\begin{align}
    &\theta_{k+1} = \theta_k + \alpha \nabla L_\theta\\
    \label{eq20}
    &L_V(\phi) = \mathbb{E} \left[ \left( \mathcal{V}(S_t) - G_t \right)^2 \right]
\end{align}
where $\alpha$ is the learning rate. $\mathcal{V}(S_t) $ is the estimated state value, and $G_t$ is the target return. Expression in~\textcolor{myblue}{\eqref{eq20}} is to minimize the discrepancy between $\mathcal{V}(S_t) $ and $ G_t $, thereby improving the accuracy of policy evaluation.  

Instead of directly estimating the value of the action-value function $\mathcal{Q}(s, a)$, the PPO algorithm relies on the state-value function $\mathcal{V}(S_t)$. This modification simplifies learning, as estimating $\mathcal{Q}(S, a)$ for all possible actions introduces high variance and computational complexity. By employing $\mathcal{V}(S_t)$, PPO efficiently derives the advantage function, as expressed in~\textcolor{myblue}{\eqref{eq21}}. This formulation enables stable policy updates while maintaining computational efficiency.
\begin{align}
    \label{eq21}
    &\mathcal{A}_t = G_t - \mathcal{V}(S_t)\\
    &G_t = \sum_{k=0}^{\infty} \gamma^k R_{t+k}
\end{align}
The interaction between agents and environment is repeated until convergence criteria (e.g., maximum steps) are met, and the optimal policy $\pi$ is obtained to control the charging and discharging of ESSs for voltage magnitude regulation using minimal active power.

\subsection{Thevenin Theorem}
\label{2.3}
\begin{figure}
    \centering
    \includegraphics[width=1.0\linewidth]{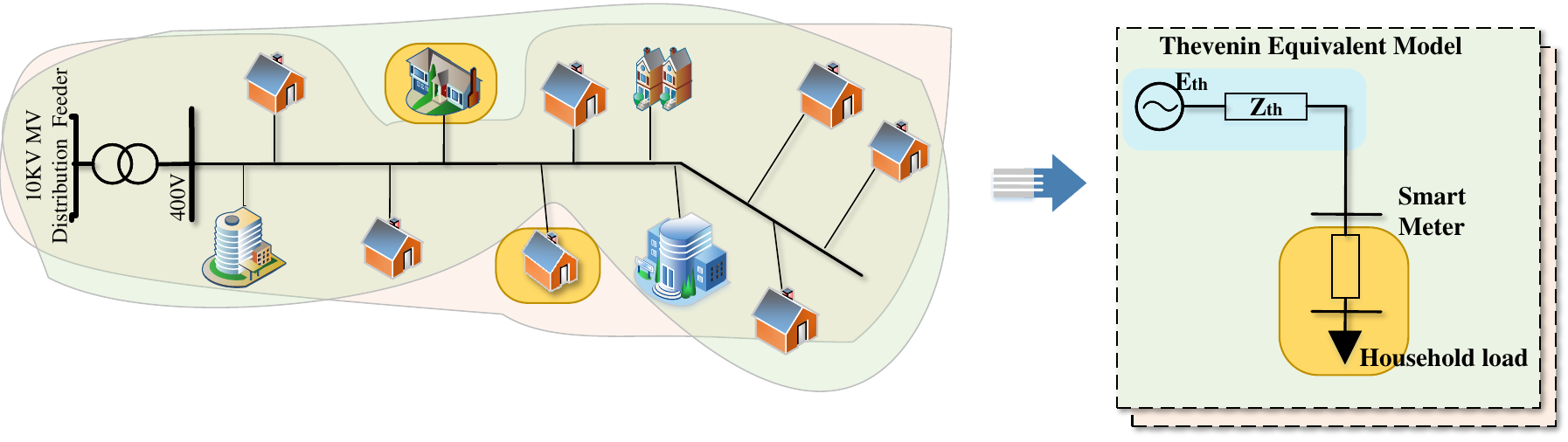}
    \caption{\textnormal{Illustrative example of distributed environment construction for smart agents in the LVDN with two ESSs.}}
    \label{fig3}
    \vspace{-0.3cm}
\end{figure}
According to the Thevenin theorem, a linear circuit comprising multiple voltage sources and impedances can be reduced to an equivalent circuit consisting of a single voltage source $E^{th}$ in series with an equivalent impedance $Z^{th}=r^{th}+jx^{th}$. This simplified representation preserves the electrical characteristics of the original network at the terminals of interest, thereby facilitating the analysis of system behavior at specific nodes \textcolor{myblue}{\cite{4275203}}. The electrical perspective observed from each SM differs, as illustrated in \textcolor{myblue}{Fig.~\ref{fig3}}.

In particular, the nodal voltage at bus $m$ at time steps $t$ and $t+1$ can be described using a Thevenin equivalent formulation, which expresses the voltage as a function of the equivalent impedance and the injected current \textcolor{myblue}{\cite{7232289}}, as shown in expressions \textcolor{myblue}{(\ref{eq23})} and \textcolor{myblue}{(\ref{eq24})}. Consequently, determining the Thevenin equivalent parameters requires two consecutive samples corresponding to distinct operating points.
\begin{align}
    \label{eq23}
    & \dot{E}_{m,t+1}^{th} = \dot{Z}_{m,t+1}^{th} \dot{I}_{m,t+1} + \dot{V}_{m,t+1} \\
    \label{eq24}
    & \dot{E}_{m,t}^{th} = \dot{Z}_{m,t}^{th} \dot{I}_{m,t} + \dot{V}_{m,t},
\end{align} 
where $\dot{I}_{m,t}$ is the current from the network to node $m$ at time $t$. When the two sampling times are very close, $\dot{E}^{th}$ and $\dot{Z}^{th}$ are assumed to be constant during this period. The approximate values of the Thevenin equivalent parameters can be inferred as follows:
\begin{align}
\label{eq25}
&\dot{Z}_m^{th} = \frac{\dot{V}_{m,t} - \dot{V}_{m,t+1}}{\dot{I}_{m,t+1} - \dot{I}_{m,t}} \\
\label{eq26}
&\dot{E}_m^{th} = \frac{\dot{V}_{m,t} \dot{I}_{m,t+1} - \dot{V}_{m,t+1} \dot{I}_{m,t}}{\dot{I}_{m,t+1} - \dot{I}_{m,t}}.
\end{align}

Therefore, by online monitoring voltage magnitudes and current values, the overall dynamics of the entire network can be observed through a node. This feature provides a reference for local control on the SM side. Assuming that the two parameters $\dot{E}^{th}$ and $\dot{Z}^{th}$ remain unchanged between two samples~\footnote{$\dot{Z}_\mathrm{th}$ is assumed constant impedance because no topological changes are expected between two samples, and no tap changes occur. $\dot{E}_\mathrm{th}$ refers to the open-circuit voltage measured or calculated across the two specific terminals of the original circuit when the load is removed.}, it is possible to calculate whether the load at the next moment will cause the voltage magnitude to exceed the limit based on these two values, which provides the possibility of pre-control at the same time.

Specifically, based on the Thevenin equivalent voltage $\dot{E}^{th}$, impedance $\dot{Z}^{th}$, and load power $P_{m}$ at the next time step, the voltage magnitude $V_{m}$ can be calculated based on expression~\textcolor{myblue}{(\ref{eq5})~\cite{7232289}}. Assuming $cos\theta$ is the power factor of node $m$ and reactive power $Q_m$ in \textcolor{myblue}{\eqref{eq50}} is $P_{m} tan\theta$. The expression in \textcolor{myblue}{(\ref{eq5})} can be re-written as in~\textcolor{myblue}{(\ref{eq50})}, which is a standard $ax^4+bx^2 + c = 0$.
\begin{align}
    & V_m^4 
  + V_m^2 \!\left[-2\!\left(r_m^{\mathrm{th}} P_m + x_m^{\mathrm{th}} Q_m\right)
  - \left(E_m^{\mathrm{th}}\right)^2 \right] \nonumber \\
  \label{eq50}
  & \hspace{1.9cm}+ \left(P_m^2 + Q_m^2\right)
  \left[\left(r_m^{\mathrm{th}}\right)^2 + \left(x_m^{\mathrm{th}}\right)^2\right] = 0.
\end{align}
The solution of expression \textcolor{myblue}{(\ref{eq50})} can be obtained as $V^2_m = \frac{-b \pm \sqrt{b^2 - 4ac}}{2a}$. In this case, due to the nature of voltage, the solution to expression~\textcolor{myblue}{(\ref{eq50})} should be $>$ 0, resulting in $V_m = \sqrt{\frac{-b + \sqrt{b^2 - 4ac}}{2a}}$. Thus, this local PF calculation enables local users to estimate voltage magnitude without relying on synchronized data collection and the centralized PF model.
\begin{figure}
\vspace{0.2cm}
    \centering
    \includegraphics[width= 0.5\textwidth]{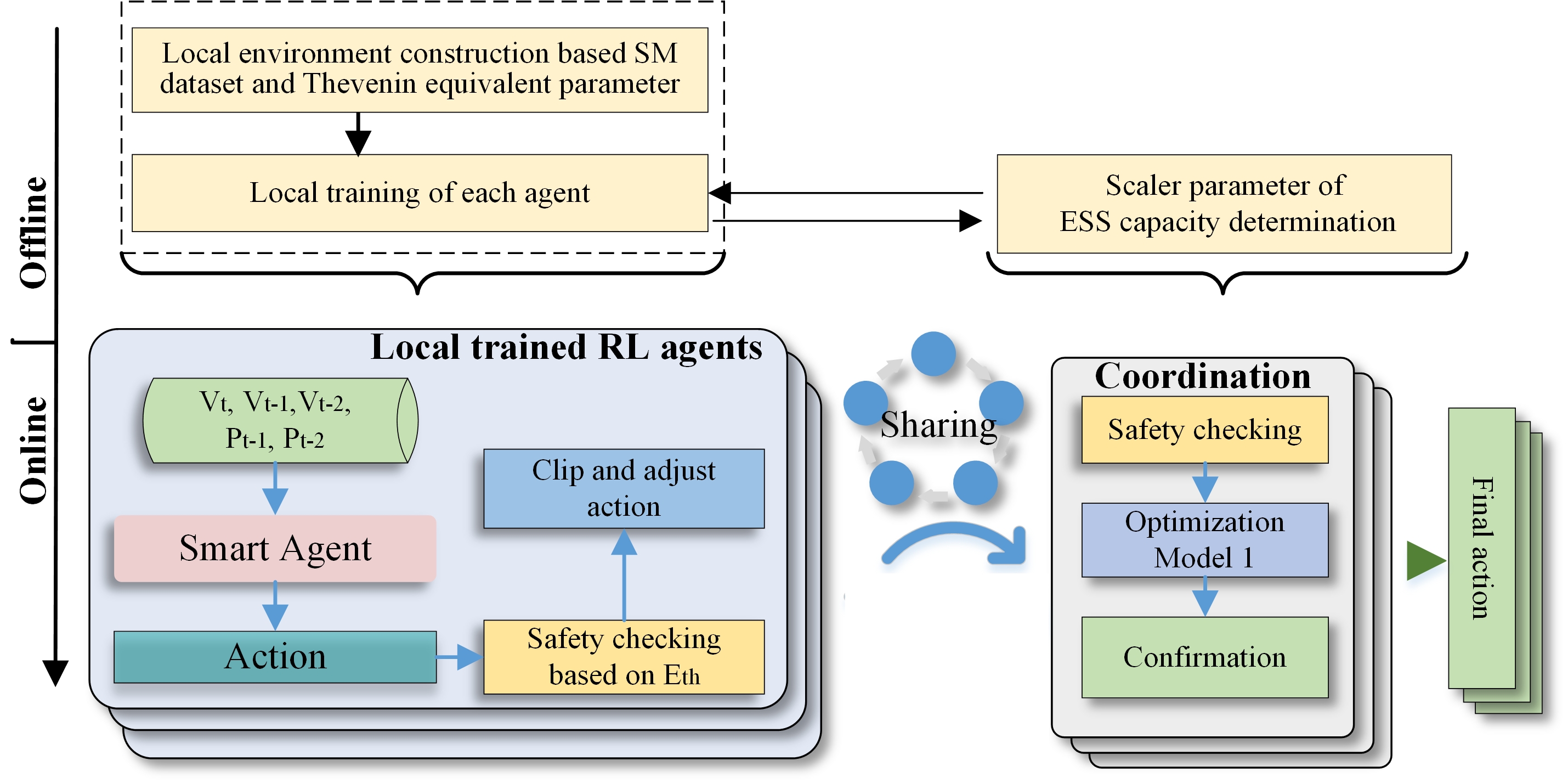}
    \vspace{-0.3cm}
    \caption{\textnormal{Coordinated-Distributed Framework for ESS-based Voltage pre-active management}}
    \label{fig2} 
    \vspace{-0.3cm}
\end{figure}

\section{Distributed RL-based Voltage Magnitude Regulation}
A distributed RL algorithm is proposed to enable RL agents deployed at households to perform voltage magnitude regulation in LVDNs without requiring centralized training and synchronicity of data collection for PF calculation. The framework of the proposed algorithm is depicted in \textcolor{myblue}{Fig.~\ref{fig2}}, which consists of independent training of local RL agents and online coordination. This paper treats household ESSs as the only dispatchable appliances. However, the application of the proposed algorithm is not limited to ESSs dispatching and can also be applied to other appliances.

\subsection{Distributed RL Agent Training}
\label{3.1}
\subsubsection{Local Voltage Magnitude Estimation}
\label{3.1.1}
The voltage magnitude estimation process described in Section~\ref{2.3} is embedded within the local environment. In this environment, the values of $E^{\mathrm{th}}$ and $Z^{\mathrm{th}}$ are dynamically updated via \textcolor{myblue}{\eqref{eq25}} and \textcolor{myblue}{\eqref{eq26}} according to the SM data sampling frequency. Given the latest values of $E^{\mathrm{th}}$ and $Z^{\mathrm{th}}$, the voltage magnitude is then estimated by solving \textcolor{myblue}{\eqref{eq30}}, which is implemented in the local environment and expressed as:

\vspace{-0.3cm}
\begin{align} 
\label{eq30}
&\hat{V}_t = 
\frac{1}{2}\{
2 (r_m^{\mathrm{th}} P_m + x_m^{\mathrm{th}} Q_m) + (E_m^{\mathrm{th}})^2
\nonumber \\
& \hspace{1cm}+[[-2 (r_m^{\mathrm{th}} P_m + x_m^{\mathrm{th}} Q_m) - (E_m^{\mathrm{th}})^2]^2 \\
& \hspace{1cm}- 4 (P_m^2 + Q_m^2)((r_m^{\mathrm{th}})^2 + (x_m^{\mathrm{th}})^2)]^{\frac{1}{2}}
\}^{\frac{1}{2}} \nonumber
\end{align}

However, the accuracy of local voltage magnitude estimation is affected by three sources of uncertainty. First, SM data are typically sampled at relatively low frequencies (e.g., every 5 or 15 minutes), which may induce error for parameters $E^{th}$ and $Z^{th}$ estimation. Second, measurement noise may further introduce inaccuracies in the estimated voltage magnitude $\hat{V}_t$. Third, when the sampled current values $\dot{I}_{m,t}$ and $\dot{I}_{m,t+1}$ are very close, the estimated voltage magnitudes via \textcolor{myblue}{\eqref{eq25}-\eqref{eq50}} fall outside the expected operational range. Thus, to ensure that the estimated voltage magnitude remains within a reasonable range (e.g., 0.95 p.u. to 1.05 p.u.), a piecewise adjustment function $\mathcal{H}(\cdot)$ is introduced to correct the estimated voltage magnitude, preventing extreme deviations caused by inaccurate parameters $E^{th}$ and $Z^{th}$. The proposed piecewise adjustment function $\mathcal{H}(\cdot)$ is formulated as follows:
\begin{align}
\label{eq31}
V_t^{\mathrm{adj}} = \mathcal{H}(\hat{V}_t)=
\begin{cases} 
    \mathcal{H}_{\mathrm{L}}(\hat{V}_t), & \hspace{0.7cm} \hat{V}_t < \tilde{V}_{\min}, \\[4pt]
    \mathcal{H}_{\mathrm{M}}(\hat{V}_t), & \hspace{0.7cm} \tilde{V}_{\min} \leq \hat{V}_t \leq \tilde{V}_{\max}, \\[4pt]
    \mathcal{H}_{\mathrm{H}}(\hat{V}_t), & \hspace{0.7cm} \hat{V}_t > \tilde{V}_{\max}.
\end{cases}
\end{align}
where $\mathcal{H}_{L}(\cdot)$, $\mathcal{H}_{M}(\cdot)$ and $\mathcal{H}_{H}(\cdot)$ are high-order polynomial fitting functions in different intervals, whose parameters are obtained on historical datasets. $\tilde{V}_{\min}$ and the $\tilde{V}_{\max}$ are the threshold values for fitting. $V_t^{\text{adj}}$ is the adjusted value of the estimated voltage magnitude $\hat{V}_t$ at time $t$.

Furthermore, the Thevenin equivalent model is a simplified model, whose sensitivity to the agent’s actions differs from that of the actual network. This discrepancy introduces estimation errors, which may lead to agent confusion (i.e., challenging to judge whether the action is good or bad). Specifically, for the voltage fluctuations, it is uncertain whether these fluctuations stem from its actions or Thevenin equivalent modelling inaccuracies. Consequently, this uncertainty may reduce the convergence of the learning process by exploring more possibilities. To address this issue, a nonlinear sensitivity function $\mathcal{E}(\cdot)$ is incorporated into the voltage magnitude estimation process in the local environment. In this work, a Transformer $ANN^0$ is employed to approximate this nonlinear relationship between actions and the sensitivity of networks to the variation of actions. The Transformer architecture is composed of three main components:

\begin{itemize}
    \item \textit{Input Embedding Layer}: The input vectors are first projected into a preset dimensional presentative, which allows the model to work in a representation space suitable for the Transformer encoder.
    
    \item \textit{Transformer Encoder}: The encoder captures long-range dependencies and complex interactions among the input features. This structure allows the model to focus adaptively on important aspects of the input for each prediction.
    
    \item \textit{Regression Head}: The encoded representation is passed through a final linear layer that outputs a scalar value, corresponding to the predicted voltage magnitude: $\tilde{V}_t$ 
\end{itemize}

The final voltage magnitude estimated based on local SM data and $V^{adj}_t$ is formulated as:
\begin{equation}
\label{eq32}
\tilde{V}_t = \mathcal{E}(a_t, P^{D}_t, V_{t-1}, V_{t-2}, P_{t-1}, P_{t-2}, E^{th}, Z^{th}, V_t^{adj})
\end{equation}

\begin{figure}
    \centering
    \includegraphics[width=0.5\textwidth,height=0.25\textwidth]{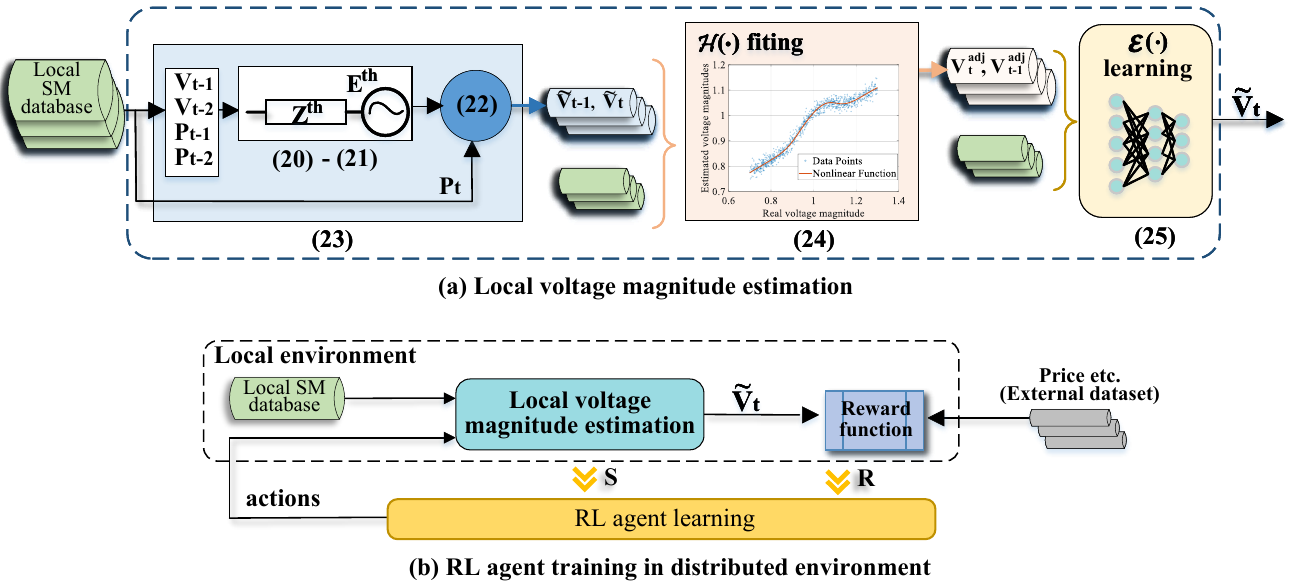}
     \caption{\textnormal{Framework of the proposed distributed RL algorithm: (a) local voltage estimation, and (b) RL agent training in the constructed local environment.}}
     \vspace{-0.2cm}
     \label{fig4}
\end{figure}
The input of the function $\mathcal{E}(\cdot)$ consists of the action $a_t$, SM data, parameters $E^{th}$ and $Z^{th}$, and the estimated voltage magnitudes $V^{adj}_t$. These inputs are carefully selected to enhance the interpretability of the learning process from a physics-informed perspective. The whole process of voltage magnitude estimation using local SM data is depicted in \textcolor{myblue}{Fig.~\ref{fig4}(a)}.

\subsubsection{Distributed Environment Construction}  
\label{3.1.2}
The SM in the household equipped with an ESS is considered an individual agent. The agent has access to local datasets, including active power and voltage magnitude measurements, while the power factor is assumed to be known. The distributed environment is constructed to estimate the voltage magnitude based on the latest SM data and the updated actions from agents, and return the corresponding state $S_t$ and reward $R_t$. In the distributed environment, each agent has one local environment, where the local voltage magnitude estimation process in Section~\ref{3.1.1} is integrated for voltage magnitude estimation using local SM data. All agents are trained independently within their respective local environments. The general training process of a single RL agent in a distributed environment is illustrated in \textcolor{myblue}{Fig.~\ref{fig4}(b)}.

The action $a_t$ represents the charging and discharging power of the ESS, modelled as variable $P^b_t$. The action space is constrained by the ESS charging/discharging limits, as formulated in expression~\textcolor{myblue}{\eqref{eq7}}.

The state vector $S_t$ at time step $t$ is defined as:  
\begin{equation}  
S_t = [P^{D}_t, SOC_t, \tilde{V}_t, V_{t-1}, V_{t-2}, P^{D}_{t-1}, P^{D}_{t-2}]
\end{equation}  
where $P^{D}_t$ represents the active power of the household at time $t$, $SOC_t$ is the state of charge of the ESS at time~$t$, and $\tilde{V}_t$ is the estimated voltage magnitudes. The inclusion of past voltage and power values $(V_{t-1}, V_{t-2}, P^D_{t-1}, P^D_{t-2})$ allows the agent to capture temporal dependencies and are used to calculate $E^{th}$ and $Z^{th}$.

The reward function $R_t$ is defined using a step function to penalise voltage magnitude violations beyond acceptable limits. It is updated as follows:  
\begin{equation}
\label{eq37}
    R_t =
    \begin{cases} 
        -w_pC^b_t\sum|P^b| - w_{\lambda}\lambda, & \quad  \tilde{V}_t > \overline{V} \text{ or } \tilde{V}_t < \underline{V} , \\
         -w_p\sum |P^b|, & \quad \underline{V} <= \tilde{V}_t <= \overline{V} .
    \end{cases}
\end{equation}
where $w_p$ and $w_{\lambda}$ are weight coefficients. $\lambda$ is a large positive penalty when there is a voltage magnitude violation in DNs. This reward function ensures that the agent prioritises actions that maintain voltage magnitude within permissible limits.

Compared to the centralized RL algorithm, the distributed RL model for voltage control faces two challenges: how to guide the agent toward convergence in scenarios with sparse positive rewards (e.g., when voltage magnitude violations occur infrequently). In most operating scenarios, the voltage magnitude in DNs remains within acceptable voltage limits. Consequently, the RL agent receives only limited guidance, regardless of whether a ramp-type or step-type reward function is used for effective convergence. Specifically, under normal operating conditions, moderate control actions do not induce voltage magnitude violations. As a result, when such data is reused during training, it provides minimal guidance for policy improvement. Besides, voltage magnitude violation mitigation by an individual user is challenging, as the voltage magnitude at that user depends not only on its own demand but also on the demand of other users connected to the same feeder.To address this challenge, relaxing factors $\underline{\varsigma}$ and $\overline{\varsigma}$ are added to the reward function. To ensure that the voltage region remains non-empty (i.e., $\overline{V} - \overline{\varsigma} > \underline{V} + \underline{\varsigma}$) and reasonable, they are selected from $[0, 0.5]$ p.u..
\begin{equation}
\label{eq38}
    R_t =
    \begin{cases} 
        -w_pC^b_t\sum|P^b_{m}| - w_{\lambda}\lambda, & \hspace{0.3cm} \tilde{V}_t > \overline{V}- \overline{\varsigma} \quad \text{or} \\
        &\hspace{0.3cm}\tilde{V}_t < \underline{V} + \underline{\varsigma}, \\
         -w_p\sum |P^b_{m}|, &  \quad \underline{V} <= \tilde{V}_t <= \overline{V}.
    \end{cases}
\end{equation}

The SOC of the ESS at time $t$, denoted as $SOC_t$, is updated according to \textcolor{myblue}{\eqref{eq9}} and \textcolor{myblue}{\eqref{eq10}}. The remaining state variables are updated sequentially as follows: 
\begin{align}
    & V_{t-2} = V_{t-1}, \\
    & V_{t-1} = V_{t}, \\
    & P^b_{t-2} = P^b_{t-1}, \\
    & P^b_{t-1} = P^b_{t}.
\end{align}

Each agent at the households is trained locally based on the local environment, which updates the state $S_t$ and rewards $R_t$. In this situation, $R_t$ is obtained by $\tilde{V}_t$ and the expression \textcolor{myblue}{\eqref{eq38}}, guiding the agent toward an approximate convergence direction. 

Nevertheless, independently trained agents encounter two main challenges when deployed in multi-agent voltage magnitude regulation systems. To maintain voltage magnitude within acceptable limits, local agents either (i) rely on redundant ESS capacity or (ii) tend to employ overly large control actions (e.g., higher charging or discharging power than expected for global voltage magnitude regulation), which is learned during the local training process. 

This arises from the coupled nature of voltage magnitude profiles in distribution networks, where the voltage magnitude at a node is affected not only by its local power injection but also by the load dynamics of adjacent nodes. Consequently, when there is no regulation strategy at neighbouring nodes, a single agent may require stronger corrective actions to ensure that the local voltage magnitude remains within operational limits. Therefore, without awareness of the actions from other independently trained RL agents, agents solely rely on local voltage magnitude conditions, typically charging when the voltage magnitude is high and discharging when it is low, leading to excessive aggregated charging/discharging for global voltage magnitude regulation. These uncoordinated and synchronous behaviours may unintentionally lead to load peak value shifting (other than mitigation) and voltage magnitude violations. 

\subsection{Online Coordination}
To mitigate excessive charging or discharging from independent RL agents. The online deployment of the proposed distributed RL algorithm is presented as Algorithm \ref{algorithm1}. First, a coordination scaler for redundant ESS capacity is introduced, which is represented by $\beta$. Action $\tilde{a}_{m}$ is the real action applied to regulate the charging and discharging of the ESS.
\begin{align}
    \label{eq399}
    & \tilde{a}_{m}  = \beta \ a_{m}\\
    & \tilde{E}^{rated}_m = \beta E^{rated}_m 
\end{align}

Second, an optimization model that allows independently trained agents to coordinate voltage magnitude regulation by dynamically adjusting actions, mitigating excessive charging or discharging. During online application, we assume that each agent shares information with its neighboring agents $\Omega$, including action $a_t$ and the estimated voltage magnitude $\tilde{V}_t$.

\begin{algorithm}[!t]
    \caption{Online deployment of locally trained RL agents}
    \label{algorithm1}    
    \KwIn{$\underline{V}, \overline{V}$, $\underline{P}^b,\overline{P}^b$, $\overline{SOC},\underline{SOC}$, $\beta$,  $ANN^0$, Trained agents}
    
    \For{each agent $m \in \mathcal{B} $}
    {Recall voltage $V_{t-1}, V_{t-2}$, loads $P_{t-1}, P_{t-2}$\\
    Action generation $a_{t} \gets$ Trained agent$(S_{t})$ \\
    Estimate $Z^{th}, E^{th}$ via $\textcolor{myblue}{\eqref{eq25}}$ and $\textcolor{myblue}{\eqref{eq26}}$\\
    $\hat{V}_t \gets$ via expression \textcolor{myblue}{\eqref{eq30}}\\
    $V^{adj}_t \gets $ $\mathcal{H}(\cdot)$ via $\textcolor{myblue}{\eqref{eq31}}$\\
    $\tilde{V}_t \gets \mathcal{E}(\cdot)$ via \textcolor{myblue}{\eqref{eq32}}\\
    \eIf{$\tilde{V}_t < V_{\min}$ \textbf{or} $\tilde{V}_t > V_{\max}$}{
    Adjust $a_{t}$ to prevent severe voltage magnitude violations \\
    }{
    Share $\tilde{V}_t$ and $a_{t}$ with neighboring agents }
    }
    $\tilde{a}_1,\cdots,\tilde{a}_1 \gets \beta$ and expression \textcolor{myblue}{\eqref{eq399}}\\
    Update action $a^{*} \gets$ sovel model $\textcolor{myblue}{\eqref{38}}-\textcolor{myblue}{\eqref{eq41}}$ locally \\
    Apply actions $a^{*}$ to the ESS locally\\
    \KwOut{Optimized actions for each agent}
    
\end{algorithm}

The objective function in \textcolor{myblue}{\eqref{38}} is designed to minimise adjustments of actions during online coordination, preserving consistency with the agent's originally planned actions to avoid disrupting the ESS's subsequent charging and discharging schedule, while ensuring global voltage magnitude regulation. The optimization model is subject to constraints, including voltage magnitude limits and bounds on action adjustments. The estimated voltage magnitude profile is represented as a vector 
$\tilde{V}_{\text{est}}=[\tilde{V}_{1}, ..., \tilde{V}_{|\Omega|}]$.
\begin{align}
\label{38}
&\min_{\mathbf{a}} \quad \sum_{m=1}^{\mathcal{B}} (a_{m,adj} - \tilde{a}_{m})^2 \\
&\textbf{s.t.} \nonumber\\
& \tilde{V}_{\text{est}} = \mathcal{F}_v(\tilde{a}_{1},...,\tilde{a}_{|\Omega|},\tilde{V}_1,...,\tilde{V}_{|\Omega|}), \hspace{1.58cm} \forall m \in \mathcal{B}\\
& \underline{V} \leq \tilde{V}_{\text{est}} \leq \overline{V}, \hspace{3.9cm} \forall m \in \mathcal{B}\\
\label{eq41}
& \underline{\text{P}}^b \leq a_{m,adj} \leq \overline{\text{P}}^b, \hspace{3.55cm} \forall m \in \mathcal{B}
\end{align}
The function $\mathcal{F}_v(\cdot)$ captures the voltage sensitivity to actions from multiple agents and is used to estimate voltage magnitudes. This estimation depends on both the local agent’s decisions and voltage magnitudes, as well as those from neighboring agents. Action adjustments are constrained by the state of charge $SOC_t$ and the charging/discharging power limit.

When active power variations are relatively small, the relationship between voltage magnitude variations and active power injections (i.e., $a_t$) could be approximated using a linear sensitivity model, as detailed in~\textcolor{myblue}{\cite{7364281}}. Therefore, a high-order polynomial function $\mathcal{F}_v(\cdot)$ is adopted, as it provides sufficient flexibility to approximate the nonlinear relationship between active power variation and voltage magnitude variations. The parameters are derived by fitting historical data. 

The dataset used to estimate the parameters of the function $\mathcal{F}_v(\cdot)$ can be extracted from historical data collected during the online operation of LVDNs with locally trained RL agents operating without a coordination strategy. The resulting action adjustments, denoted as $a_{m, \text{adj}}$ are then transferred to the ESS for execution. Thus, the proposed algorithm offers a solution for both local training and coordinated voltage magnitude regulation, ensuring improved steady-state performance and convergence efficiency.

\begin{figure}
\vspace{0.2cm}
    \centering
    \includegraphics[width=0.43\textwidth,height=0.15\textwidth]{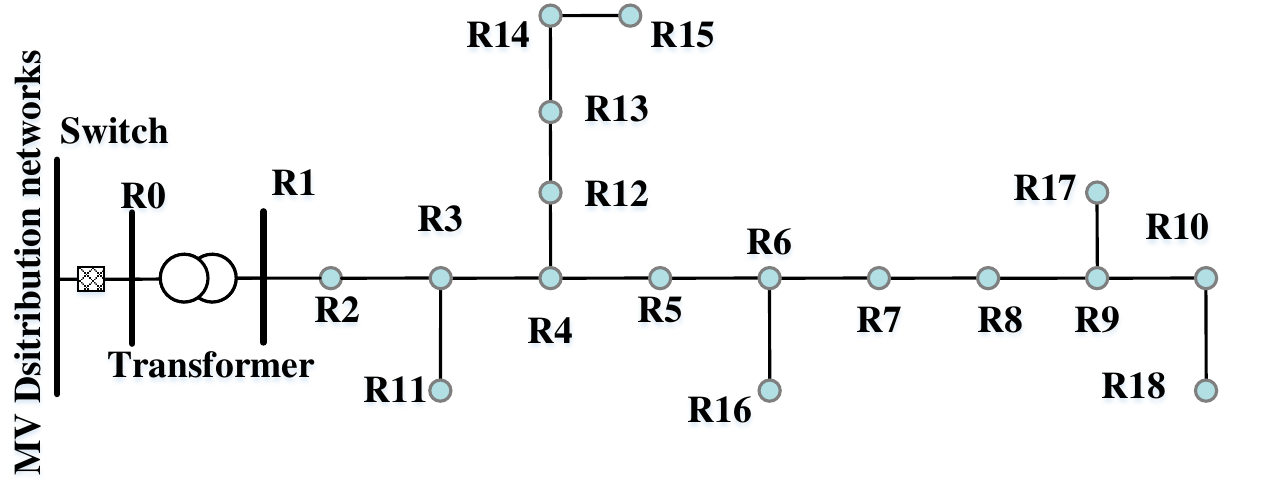}
    \vspace{-0.1cm}
     \caption{\textnormal{Topology of the 18-bus CIGRE LVND.}}
     \label{fig6}
     \vspace{-0.5cm}
\end{figure}
\section{Case Study}
The CIGRE low-voltage network is utilised as the test case, where each node is connected to specific loads, as illustrated in~\textcolor{myblue}{Fig.~\ref{fig6}}. The corresponding load profiles are sourced from \textcolor{myblue}{\cite{hou2024distflow}}. The time resolution of the datasets is 15 minutes. The PF results obtained using the Pandapower-based computation serve as the benchmark for centralized calculations. 

The piecewise function $\mathcal{H}(\cdot)$ is composed of three segments, each modelled using a fifth-order polynomial. These segments, defined over the intervals [0, 390 V], [390 V, 400 V], and [400 V, $\infty$], are determined through cross-validation to capture the nonlinear voltage characteristics. The order of the polynomial functions $\mathcal{H}_L(\cdot)$, $\mathcal{H}_M(\cdot)$ and $\mathcal{H}_H(\cdot)$ is set to 3.  Within the RL algorithm, a Transformer is integrated in the local environment for local voltage magnitude estimation. The architecture includes two Transformer encoder layers, each consisting of multi-head self-attention, a two-layer multilayer perceptron, and Layer Normalisation. The learning rate is set to $1 \times 10^{-3}$. 

In the RL algorithm, the learning rate is set to $ 1 \times 10^{-3} $, and the discount factor is defined as $ \gamma = 0.5 $. ESSs are installed at nodes R9, R14, and R16. The maximum rated power is set to 60~kW, the capacity is set to 3~MWh, and the coordination scaler (i.e., $\beta$ in \textcolor{myblue}{\eqref{eq399}}) is selected from [0,1]. The pre-set voltage magnitude range is specified as $[0.95, 1.05]$~p.u. (i.e., a commonly used interval in DNs), while the state-of-charge $ SOC_t $ is constrained within $[10\%, 90\%]$. In the coordination layer, the order of the polynomial function $\mathcal{F}_v(\cdot)$ is set to 4.

\subsection{Voltage Magnitude Estimation}
\begin{figure}
    \centering
    \includegraphics[width=0.43\textwidth,height=0.6\textwidth]{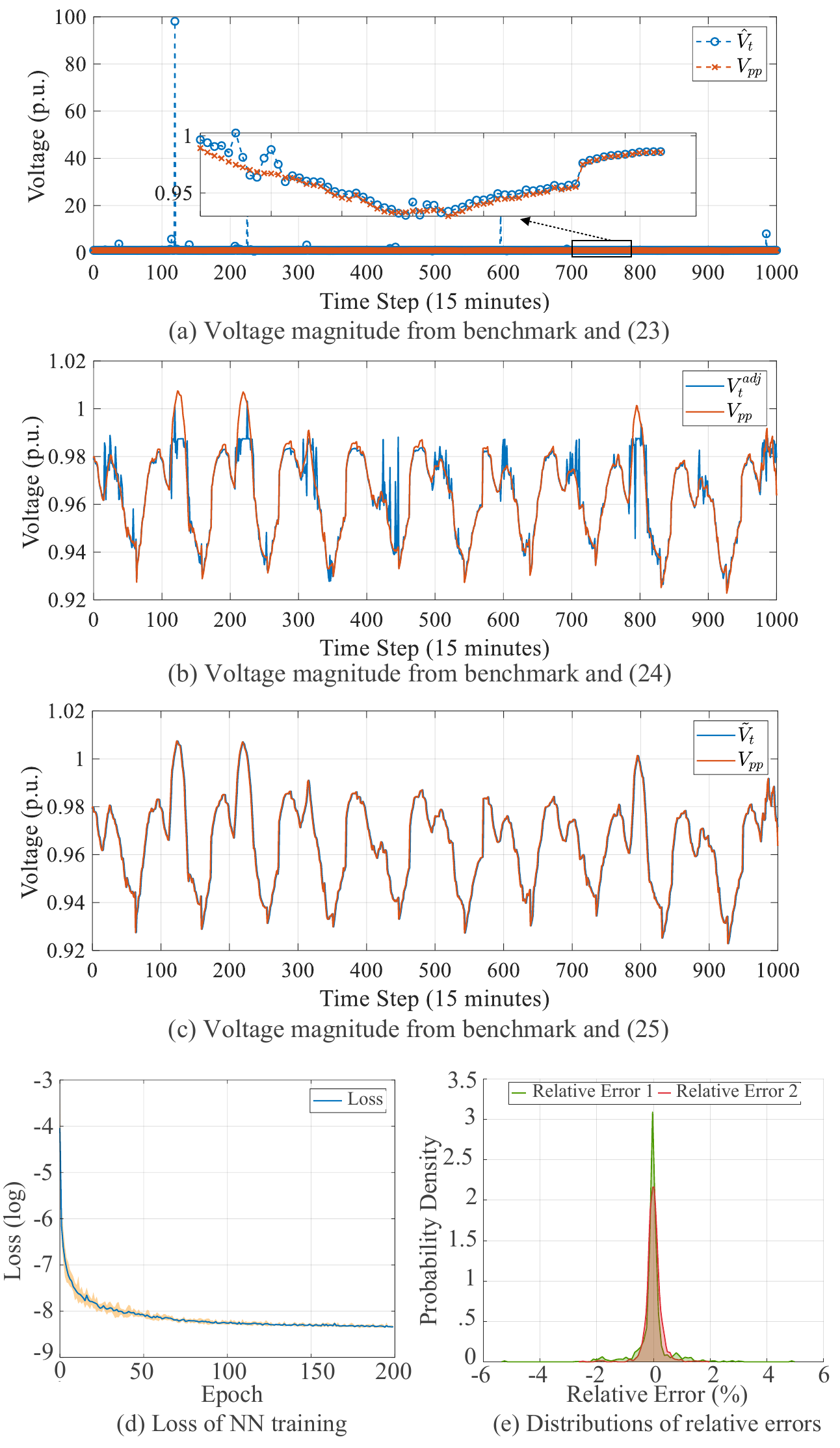}
     \caption{\textnormal{Local voltage magnitude estimation and errors distribution: (a)-(c) estimated voltage magnitude, (d) loss of the ANNs training, and (e) distribution of relative errors.}}
     \vspace{-0.2cm}
     \label{fig7}
\end{figure}

Based on the voltage magnitude estimation process in Section~\ref{3.1.1}, the estimated voltage magnitude at node $R10$ from each sub-step are depicted in \textcolor{myblue}{Fig.~\ref{fig7}}. $V_{PP}$ represents the benchmark voltage magnitude obtained from Pandapower (i.e., taken as benchmark). The top figure in \textcolor{myblue}{Fig.~\ref{fig7}(a)} depicts the estimated voltage magnitude $\hat{V}_t$ only using dynamic $E^{th}$ and $Z^{th}$, where there are multiple significant errors (e.g., 100 p.u. at time step 100) induced by the abnormal value of parameters $E^{th}$ and $Z^{th}$. The loss of Transformer training is shown in \textcolor{myblue}{Fig.~\ref{fig7}(b)}. As shown in \textcolor{myblue}{Fig.~\ref{fig7}(c)}, the relative error between $V_{pp}$ and $V^{\text{adj}}_t$ (denoted as Relative Error 1) and the relative error between $V_{pp}$ and $\tilde{V}_t$ (denoted as Relative Error 2) exhibit similar trends.

Based on the first row in \textcolor{myblue}{Fig.~\ref{fig7}(a)}, excluding significant errors, the relative error between the estimated voltage magnitude $\hat{V}_t$ and the true value $V_{PP}$ ranges from 0.2\% to 2500\%. As shown in the second row in \textcolor{myblue}{Fig.~\ref{fig7}(a)}, given the piecewise function $\mathcal{H}(\cdot)$ (i.e., 3 3-order functions), the abnormal voltage magnitude is corrected into the reasonable region (i.e., 0.9-1.05 p.u.). Relative Error 1 is generally limited to a range of approximately $\pm 4\%$. This demonstrates that the proposed piecewise function compensates for voltage magnitude estimation errors arising from Thevenin parameter equivalence, bringing the estimated voltage magnitude closer to true values. Nevertheless, the adjusted voltage magnitude exhibits abrupt high-frequency peaks, resulting from the inability of the piecewise function to cover all scenarios due to parameter inaccuracies. These abrupt high-frequency peaks in the estimated voltage magnitude prevent the agents from accurately distinguishing whether the voltage magnitude changes are caused by their current actions or by inherent estimation errors, thereby affecting convergence. In addition, although the voltage magnitude becomes more accurate after modification by the piecewise function, initial results indicate that the adjusted voltage magnitude still deviates from the true value when an action is applied. This is because the simplified Thevenin equivalent and the piecewise function cannot fully imitate the sensitivity of the node to the action. 

The voltage magnitude modified by the trained Transformer is shown in the third row in \textcolor{myblue}{Fig.~\ref{fig7}(a)}. The Relative Error 2 spans a broader range, typically within $\pm 2\% $.  \textcolor{myblue}{Fig.~\ref{fig7}} depicts that the further adjusted voltage magnitude $\tilde{V}_t$ not only aligns more closely with the true voltage magnitude $V_{pp} $, but also effectively mitigate high-frequency variations. 

\begin{figure}
    \centering
    \includegraphics[width=0.49\textwidth,height=0.2\textwidth]{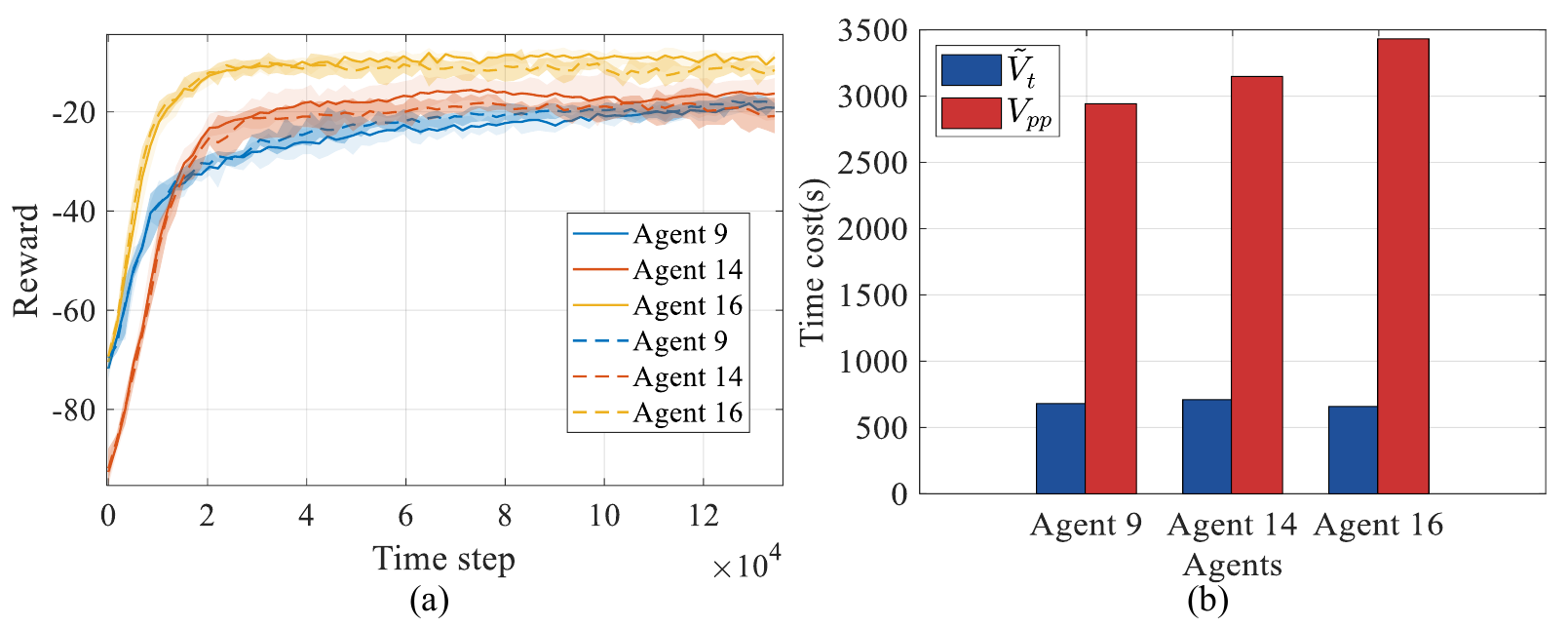}
    \vspace{-0.4cm}
    \caption{\textnormal{Comparison of rewards in (a): dashed and solid curves represent local training using the centralized PF model and the proposed distributed model, respectively; time cost in (b): red and blue bars correspond to RL training with the same setting.}}
    \vspace{-0.2cm}
     \label{fig8}
\end{figure}

\subsection{Local Offline Training}
This section provides a comparative assessment of convergence characteristics and computational efficiency for agent training under two distinct environmental configurations: the centralized PF model and an approximate voltage magnitude estimation approach, as described in Section~\ref{3.1.1}. The three ESSs are installed at nodes R9, R14, and R16, and the agents at the corresponding nodes are donated as Agent 9, Agent 14, and Agent 16, respectively. The simulation results are depicted in \textcolor{myblue}{Fig.~\ref{fig8}}.

In \textcolor{myblue}{Fig.~\ref{fig8}(a)}, the dashed curves represent the evolution of cumulative rewards during the RL agent training process using the centralized PF model, whereas the solid curve corresponds to cumulative rewards under the distributed voltage magnitude estimation approach. Despite the presence of estimation errors inherent to the distributed approach, reflected in a slight deviation in the final convergence value, the overall learning dynamics exhibit comparable convergence trends. The final reward deviation across the three training cases is less than 10, corresponding to an approximately 30\% relative error. This observation indicates that the distributed voltage magnitude estimation approach maintains sufficient fidelity to enable policy learning by the agent.

\textcolor{myblue}{Fig.~\ref{fig8}(b)} further reveals a substantial difference in training duration between the two models. Specifically, the distributed approach achieves convergence in approximately one-third of the time required by the centralized approach. All training procedures were executed on an identical computing infrastructure, thereby eliminating hardware-induced variability and ensuring comparability. It is should be emphasized that the evaluation excludes the latency associated with SM data acquisition in the centralized configuration. In practical scenarios, the process of aggregating distributed SM data is likely to introduce additional delays and computational burdens. By contrast, the decentralized approach relies on only locally available SM data, thereby obviating the need for external data transmission. This not only reduces the training time but also mitigates the privacy risks associated with centralized data collection and communication. The experimental results underscore that the proposed distributed approach constitutes a computationally efficient and privacy-preserving alternative to centralized training algorithms. Despite its simplified formulation, it supports effective agent training with minimal degradation in performance, thereby demonstrating its potential for practical deployment in DNs.

\begin{figure}
    \centering
    \includegraphics[width=0.42\textwidth,height=0.29\textwidth]{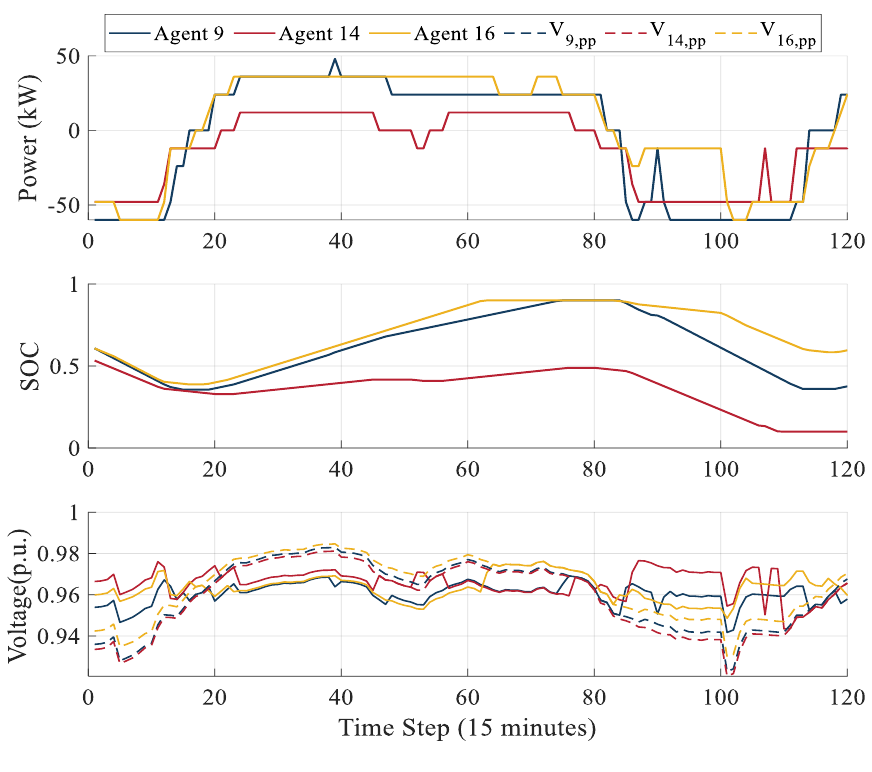}
    \vspace{-0.1cm}
     \caption{\textnormal{The power, $SOC$ of the ESS, and voltage magnitude at nodes R9, R14, and R16 with the locally trained agent: dashed lines represent that from centralized model without control, while solid lines correspond to that from the proposed approach.}}
     \vspace{-0.3cm}
     \label{fig9}
\end{figure}

During the online implementation, locally trained agents operate autonomously to regulate node voltage magnitudes without relying on centralized or coordinated control strategies. In this setup, each RL agent is deployed sequentially within DNs and evaluated independently. The charging and discharging power of the ESSs, their corresponding SOC, and the voltage magnitude at the associated bus are depicted in \textcolor{myblue}{Fig.~\ref{fig9}}. Dashed lines represent the benchmark voltage magnitudes (i.e., $V_{9, pp}$, $V_{14, pp}$, and $V_{16, pp}$) at nodes, obtained from the centralized model without any control strategy. Generally, the uncontrolled voltage profiles tend to violate the lower voltage magnitude limits more frequently than the upper limits. This is primarily attributed to high residential demand during certain periods. In response, the trained RL agents learn to discharge the ESSs to increase voltage magnitude, as evidenced by the $SOC$ trajectories. Besides, the RL agents are also capable of preserving sufficient energy within the ESSs to ensure future discharging capability when voltage magnitude support is needed. As a result, in periods of low demand or when voltage magnitudes are within acceptable ranges, the agents proactively charge the ESS, which may lead to a slight decrease in voltage magnitude. Overall,  compared to uncontrolled scenarios, the voltage profiles regulated by the agents exhibit reduced fluctuations and fewer violations of voltage limits. This indicates that locally trained RL agents, even when acting independently, are effective in voltage regulation and supporting the safe operation of the DNs. Although the agent mitigates voltage violations, the voltage magnitude at node 9 exhibits periods with $V_9<0.95$ because the ESS discharging is constrained by its maximum rated power limit. 

\begin{figure}
    \centering
    \includegraphics[width=0.4\textwidth,height=0.4\textwidth]{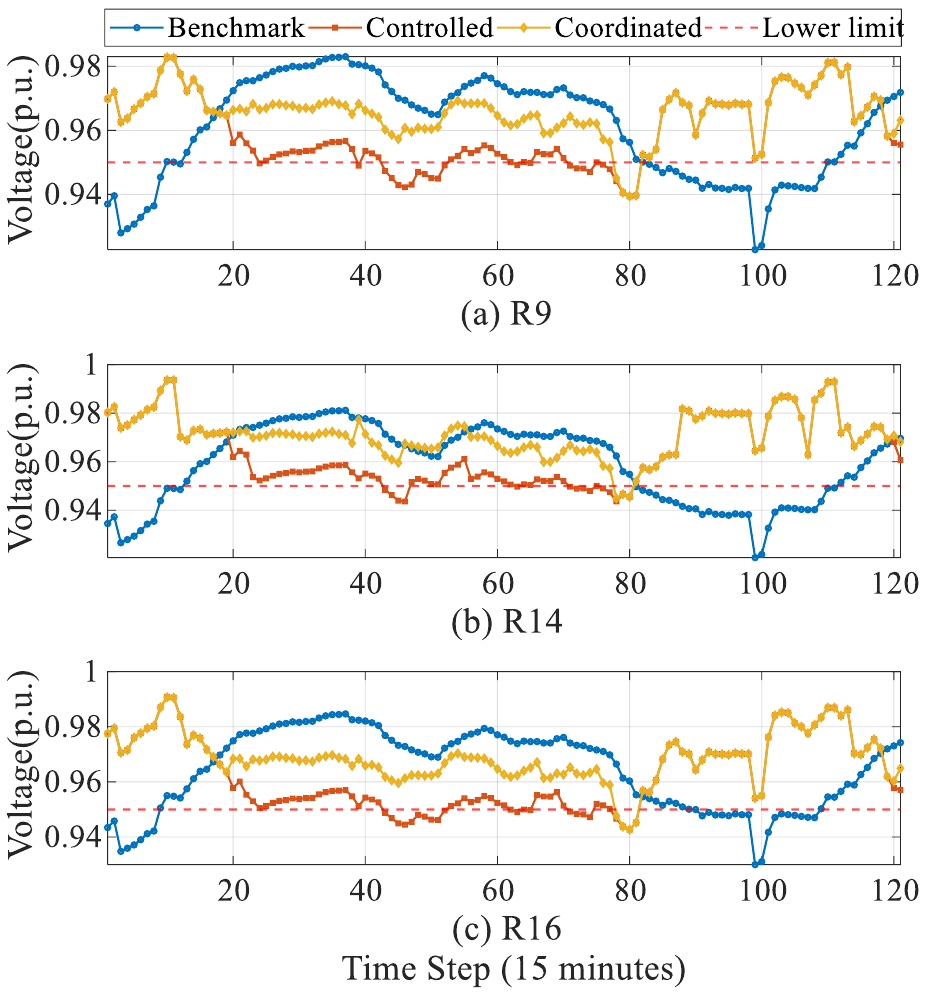}
     \caption{\textnormal{Voltage magnitude at node R9 (a), R14 (b), and R15 (c) under three conditions: no regulation (Benchmark), with individual agents (Controlled), and with coordinated agents (Controlled).}}
     \label{fig10}
     \vspace{-0.4cm}
\end{figure}

\subsection{Coordination of Agents}
This section assesses the feasibility of the proposed coordination strategy and analyzes the influence of the scaling factor $\beta$ and optimization models on overall coordination performance. In the initial scenario, three agents (i.e., located at nodes R9, R14, and R16) share information with one another. To demonstrate the scalability of the proposed coordination approach, this setup is extended by introducing additional agents at nodes R7, R15, and R18, which are neighbours of R9, R14, and R16, respectively. In this expanded scenario, each agent shares information only with its neighbours. The raw voltage magnitudes represent the baseline case without any voltage magnitude regulation. The controlled voltage magnitudes are obtained from agents trained independently, while the coordinated voltage magnitudes are obtained under the case in which independently trained agents operate under the proposed coordination strategy.

\begin{figure}
    \centering
    \includegraphics[width=0.48\textwidth,height=0.22\textwidth]{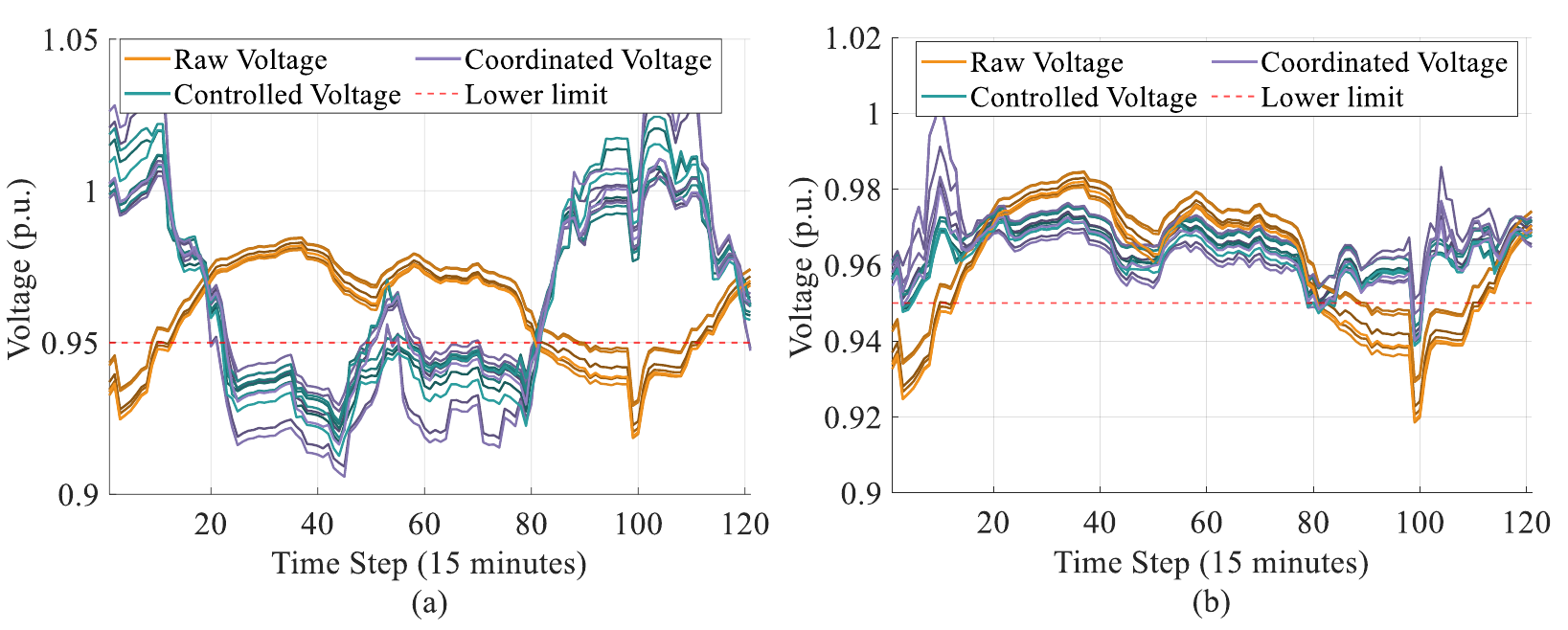}
    \vspace{-0.4cm}
     \caption{\textnormal{Voltage magnitudes at six nodes under coordinated control: (a) with a scaling factor of 1.0 (i.e., relying solely on optimization models), and (b) with a scaling factor $\beta$ of 0.2.}}
     \label{fig11}
     \vspace{-0.2cm}
\end{figure}

First, the coordination scaler $\beta$ is set as $1$, i.e., the coordination strategy relies solely on the optimisation model. The results are depicted in \textcolor{myblue}{Fig.~\ref{fig10}} and \textcolor{myblue}{Fig.~\ref{fig11}(a)}. As expected, when multiple independently trained agents are simultaneously deployed for voltage magnitude regulation, their independent decisions may lead to excessive and uncoordinated charging or discharging of multiple ESSs, shown as the deep orange curves in \textcolor{myblue}{Fig.~\ref{fig10}} and light green curves in \textcolor{myblue}{Fig.~\ref{fig11}(a)}. This behaviour results in local voltage magnitudes exceeding operational limits (e.g., time steps 40-60), as each agent is trained solely on its local voltage profile, without considering system-wide voltage magnitude interactions. This phenomenon is verified and illustrated when individually trained agents at nodes R9 to R16 are applied in parallel, a simultaneous charging event causes the voltage magnitude to drop below the lower operational threshold of 0.95 p.u.. For coordination, agents exchange both their intended actions and estimated voltage magnitude. Using this shared information, each agent adjusts individual actions to ensure coordinated behaviour, thereby mitigating adverse system-level effects. The effectiveness of this approach is demonstrated by the yellow curve in \textcolor{myblue}{Fig.~\ref{fig10}}, where the adjusted control strategy significantly reduces voltage magnitude violations. However, the purple curves in \textcolor{myblue}{Fig.~\ref{fig11}(a)} shows that the optimisation model fails to improve the coordination. This is because the adjusted space (i.e., constraint~\textcolor{myblue}{\eqref{eq41}}) does not support sufficient coordination, which is verified as follows.

Moreover, the coordination scaler $\beta$ is decreased from $1$ to $0.1$, and the performance of the coordination strategy is analysed. When $\beta$ is larger than about 0.5, the coordination strategy in the network with three agents helps DSO guarantee the voltage magnitudes in the safe region, while it is smaller than 0.5, the coordination fails since there is not enough adjustment region for each agent. However, in the networks with 6 agents, when the $\beta$ is smaller than about 0.4, the voltage magnitudes are controlled into the safe region, illustrated in \textcolor{myblue}{Fig.~\ref{fig11}(b)}. Conversely, when $\beta$ is larger than 0.4, the agents fail to maintain the voltage magnitude within the safe region. Thus, the results reveal that a large $\beta$ is suitable for coordination with more independently trained agents since the aggregated synchronous charging or discharging power is larger.

The minimal residual violations observed are primarily due to the physical constraints of the ESSs, such as the low available energy and limited rated power, which may restrict the extent to which the collaborative mechanism can correct voltage deviations. Nevertheless, the proposed algorithm effectively enhances steady-state performance and demonstrates the potential of coordinated multi-agent control in distributed voltage magnitude regulation scenarios. On the other hand, compared with the optimization model, using a scaling factor $\beta$ to achieve coordination among independently trained RL agents is more convenient. Once the scaling factor $\beta$ is determined, no interaction is required during operation, thereby reducing communication overhead. Moreover, the scaling factor can be validated among different agents under data-protection mechanisms (e.g., zero-knowledge proofs algorithm), which further lowers the risk of information leakage \textcolor{myblue}{\cite{liu2025model}}.

\section{Conclusion}
\label{conclusion}
This paper presents a distributed reinforcement learning algorithm for voltage magnitude regulation in distribution networks, focusing on decentralized approach and privacy preservation. By deploying a local environment and locally trained agents, the proposed method achieves effective voltage magnitude regulation using only local SM data, thereby minimizing the need for real-time smart meter data sharing with distribution system operators. Simulation results demonstrate that while voltage magnitude estimation using Thevenin equivalent parameters yields relatively large errors under some scenarios, the integration of a piecewise function and a neural network-based sensitivity analysis significantly enhances the accuracy of voltage magnitude estimation from local observations. Furthermore, the proposed coordination strategy enables real-time coordination using the exchange of minimal information from agents. This coordination effectively mitigates the risk of simultaneous overcharging or discharging by multiple ESSs, ensuring effective voltage regulation and reliable system operation in decentralized environments. Future work will concentrate on advancing the scalability of the proposed coordination strategy by investigating lightweight coordination mechanisms, hierarchical schemes, and adaptive scaling techniques that support deployment in large scale multi-agent systems.

\printcredits
\vspace{-0.3cm}

\bibliographystyle{elsarticle-num}
\bibliography{ref}

\end{document}